\documentclass[a4paper,11pt]{article}
\pdfoutput=1
\usepackage{jheppub}
\usepackage{amsmath,amssymb,bm,graphicx,bbold,epsf,colordvi}
\usepackage{lipsum}
\usepackage{soul}
\usepackage{braket}
\usepackage{bm}
\allowdisplaybreaks 
\usepackage{color}
\usepackage[dvipsnames]{xcolor}

\usepackage{mathrsfs}
\allowdisplaybreaks
\usepackage[abs]{overpic}
\usepackage[export]{adjustbox}
\usepackage{bm}
\usepackage{scalerel}
\usepackage{accents}
\newcommand\thickbar[1]{\accentset{\rule{.4em}{.8pt}}{#1}}

\newcommand{\bef}{\begin{figure}[hbt]\centering}
\newcommand{\eef}{\end{figure}}
\usepackage{mathtools}
\usepackage{subfigure}
\usepackage{booktabs}
\usepackage{graphicx}
\graphicspath{ {./figure/} }
\usepackage{comment}
\usepackage{lipsum}

\newcommand{\beq}{\begin{equation}}
\newcommand{\eeq}{\end{equation}}
\def\bea#1\eea{\begin{align}#1\end{align}}

\def \be  {\begin{equation}}
\def \ee  {\end{equation}}
\def \ba  {\begin{eqnarray}}
\def \ea  {\end{eqnarray}}

\newcommand{\nn}{\nonumber}

\allowdisplaybreaks

\makeatletter
\def\@fpheader{~}
\makeatother

\usepackage[Q=yes,pverb-linebreak=no]{examplep}

\title{QCD evolution of the gluon Sivers function in heavy flavor dijet production at the Electron-Ion Collider}

\author[a,b,c]{Zhong-Bo Kang}
\author[a,b]{, Jared Reiten}
\author[a,b,c,d,e]{, Ding Yu Shao}
\author[a,b]{, and John Terry}

\affiliation[a]{Department of Physics and Astronomy, University of California, Los Angeles, CA 90095, USA}
\affiliation[b]{Mani L. Bhaumik Institute for Theoretical Physics, University of California, Los Angeles, CA 90095, USA}
\affiliation[c]{Center for Frontiers in Nuclear Science, Stony Brook University, Stony Brook, NY 11794, USA}
\affiliation[d]{Department of Physics and Center for Field Theory and Particle Physics, Fudan University, Shanghai, 200433, China}
\affiliation[e]{Key Laboratory of Nuclear Physics and Ion-beam Application (MOE), Fudan University, Shanghai, 200433, China}

\emailAdd{zkang@g.ucla.edu, jdreiten@physics.ucla.edu, dingyu.shao@cern.ch, johndterry@physics.ucla.edu}

\abstract
{Using Soft-Collinear Effective Theory, we develop the transverse-momentum-dependent factorization formalism for heavy flavor dijet production in polarized-proton-electron collisions. We consider heavy flavor mass corrections in the collinear-soft and jet functions, as well as the associated evolution equations. Using this formalism, we generate a prediction for the gluon Sivers asymmetry for charm and bottom dijet production at the future Electron-Ion Collider. Furthermore, we compare theoretical predictions with and without the inclusion of finite quark masses. We find that the heavy flavor mass effects can give sizable corrections to the predicted asymmetry.
}

\begin{document}
\maketitle

\section{Introduction}
In recent years, one of the most important forefronts of hadron physics has been the exploration of the three-dimensional (3D) partonic structure of nucleons in momentum space. Such 3D information is encoded in the so-called transverse-momentum-dependent parton distribution functions (TMD PDFs), which can further inform us about the confined motion of partons in the nucleon, as well as the correlation between their spins, momenta, and the spin of the nucleon~\cite{Accardi:2012qut}. Thanks to semi-inclusive deep inelastic scattering (SIDIS), a great deal of progress has been made in probing and extracting the TMD PDFs of quarks--however, information regarding those of gluons is still largely unknown experimentally. Exploring and measuring gluon TMD PDFs is one of the primary goals for the future Electron Ion Collider (EIC). 

Among the gluon TMD PDFs, the so-called gluon Sivers function is regarded as one of the ``golden measurements" at the future EIC~\cite{Accardi:2012qut}. The gluon Sivers function encapsulates the quantum correlation between the gluon's transverse momentum inside the proton and the spin of the proton, thus providing 3D imaging of the gluon's motion. Quite a few processes have been proposed to probe the gluon Sivers function at the EIC, including heavy quark pair production~\cite{Boer:2016fqd}, heavy quarkonium production~\cite{Yuan:2008vn,Mukherjee:2016qxa,Rajesh:2018qks,Bacchetta:2018ivt,Boer:2020bbd}, and quarkonium-jet production~\cite{DAlesio:2019qpk}, as well as back-to-back dihadron and dijet production~\cite{Boer:2011fh}. The feasibility of measuring the gluon Sivers function in the above scenarios has been studied in~\cite{Zheng:2018ssm}, where the authors use the PYTHIA event generator \cite{Sjostrand:2006za} and the reweighting method of \cite{Airapetian:2010ac} to investigate the spin asymmetry. They conclude that dijet production is the most promising channel for probing gluon Sivers effects, where the selection of a sufficiently small-$x$ value suppresses the contribution of the quark channel and the corresponding quark Sivers function. In this paper, we discuss spin asymmetry in the process of heavy flavor (HF) dijet production, where the contribution of the quark Sivers function is further suppressed compared to that of the light flavor dijet case. 

An intriguing feature common to both quark and gluon Sivers functions is that they depend non-trivially on the processes in which they are probed.
A well-known example of the process-dependence of the quark Sivers function is its sign change between SIDIS and Drell-Yan processes~\cite{Brodsky:2002cx,Collins:2002kn,Boer:2003cm}. Similarly, it has been demonstrated that the gluon Sivers function for the process of back-to-back diphoton production in $p+p$ collisions, $p^\uparrow p \to \gamma \gamma X$, carries a sign opposite to that of dijet production in $e+p$ collisions, $e p^{\uparrow} \rightarrow e^{\prime} j j X$: $f_{1 T,g}^{\perp \left[e p^{\uparrow} \rightarrow e^{\prime} j j X\right]}\left(x, k_{T}\right)=-f_{1 T,g}^{\perp \left[p^{\uparrow} p \rightarrow \gamma \gamma X\right]}\left(x, k_{T}\right)$~\cite{Boer:2016fqd}. In~\cite{Buffing:2013kca},  it was demonstrated that the gluon Sivers function in any process can be expressed in terms of two ``universal'' functions with calculable color coefficients for each partonic subprocess. We briefly discuss such a process-dependence for HF dijet production below. For a comprehensive review on gluon TMD PDFs, see~\cite{Boer:2015vso,Arbuzov:2020cqg}. 

So far, studies of the gluon Sivers function at the EIC are mostly performed within the leading-order (LO) parton model, without considering the impact of QCD evolution. The effects of resummation for back-to-back light flavor dijet production in the unpolarized DIS process have been investigated in \cite{Banfi:2008qs}, where the authors apply the $p_T$-weighted recombination scheme \cite{Ellis:1993tq} in defining the jet axis to avoid the theoretical complexity arising from non-global logarithm (NGL) resummation \cite{Dasgupta:2001sh}. A similar idea is used to study single inclusive jet production in the Breit frame at the EIC in \cite{Gutierrez-Reyes:2018qez,Gutierrez-Reyes:2019vbx}. Recently, following the same Soft-Collinear Effective Theory (SCET) framework utilized in \cite{Buffing:2018ggv,Liu:2018trl,Liu:2020dct,Chien:2019gyf,Kang:2020xez}, the TMD factorization formula for light flavor dijet production at the EIC has been derived \cite{delCastillo:2020omr}, where the azimuthal-angle-dependent soft function, describing the interaction between two final-state jets through the exchange of low-energy gluons, is analytically calculated at one-loop order. For HF jet production in the kinematic region of comparable jet and heavy quark masses, a new effective theory framework is needed. In this work, we provide such a framework and derive the TMD factorization formula. 

The remainder of this paper is organized as follows. In Sec.~\ref{sec:fac}, we detail the factorization framework required to carry out resummation in the back-to-back region where the transverse momentum imbalance of the HF dijet is small. In Sec.~\ref{sec:num}, we present numerical results for charm and bottom dijet production in both unpolarized and transversely-polarized-proton-electron scattering. We summarize our findings and give an outlook for future investigations in Sec.~\ref{sec:concl}.

\section{Factorization and resummation formula}\label{sec:fac}
In this section, we start with the kinematics for HF dijet production in $e+p$ collisions. We then provide the TMD factorization formalism with explicit expressions for all the relevant factorized ingredients. 
 
\begin{figure}[htb]
    \centering
    \vspace{-2cm}
    \begin{overpic}[width=1.05\linewidth]{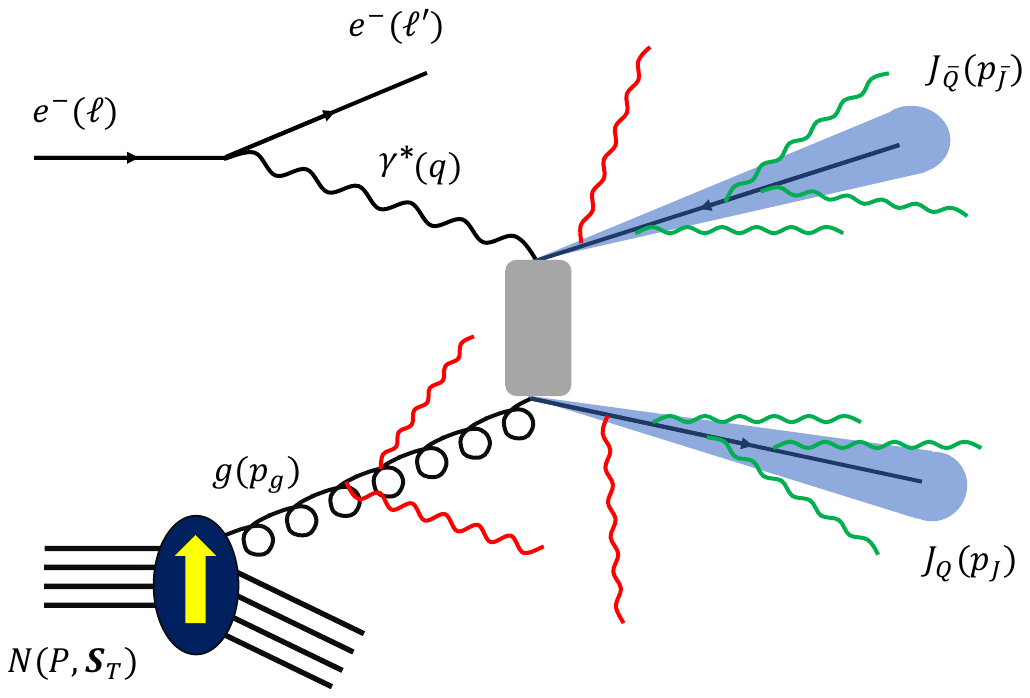}
    \end{overpic}
    \vspace{-2.5cm}
    \caption{HF dijet production in electron-proton collisions, as stated in Eq.~\eqref{eq:pro}. 
    }
    \label{fig:Qjet-pro}
\end{figure}
\subsection{Kinematics} 
As shown in Fig.~\ref{fig:Qjet-pro}, we consider HF dijet production in the polarized-proton-electron scattering process
\begin{align}\label{eq:pro}
    e(\ell) + N(P, \bm{S}_T) \to e(\ell') + J_{\mathcal{Q}}(p_J) + J_{\bar{\mathcal{Q}}}( p_{ \thickbar{J}}) + X\,,
\end{align}
where $\bm{S}_T$ is the transverse spin of the polarized proton with momentum $P$ and $\ell$ ($\ell'$) is the momentum of the incoming (outgoing) electron. At LO, HF dijets are produced via the $\gamma^*g\to \mathcal{Q}\bar{\mathcal{Q}}$ process. The HF quark $\mathcal{Q}$ and antiquark $\bar{\mathcal{Q}}$ initiate the observed HF jets $J_{\mathcal{Q}}$ and $J_{\bar{\mathcal{Q}}}$ with momentum $p_J$ and $p_{\thickbar{J}}$, respectively. In this paper, we choose to work in the Breit frame so that both the virtual photon (with momentum $q = \ell - \ell'$) and the beam proton scatter along the $z$-axis. For convenience, we define the following variables commonly used in DIS,
\begin{align}
    Q^{2}=-q^{2}\,, \quad x_{B}=\frac{Q^{2}}{2 P \cdot q}\,, \quad y=\frac{P \cdot q}{P \cdot \ell}\,. 
\end{align}
We may further note that $Q^2=x_B\, y\, S_{\ell P}$, where $S_{\ell P}=(\ell+P)^2$ denotes the electron-proton center-of-mass energy. In a fashion analogous to SIDIS, we also define the kinematic variable $z = P\cdot p_J/P\cdot q$, which gives the momentum fraction of the photon carried by the jet $J_{\mathcal{Q}}$. At LO, the four-momenta of the incoming and outgoing particles are expressed as
\begin{align}
    & q^\mu = \frac{Q}{2} (n^\mu - \bar n^{\mu})\,,\quad P^\mu = \frac{Q}{x_B} \frac{\bar n^{\mu}}{2}\,, \notag \\
    & \ell^\mu = \frac{Q}{y} \frac{n^{\mu}}{2} + Q \frac{1-y}{y} \frac{\bar n^{\mu}}{2} + \ell_{t}^\mu\,,\quad \ell'^\mu = Q \frac{1-y}{y} \frac{n^{\mu}}{2} + \frac{Q}{y} \frac{\bar n^{\mu}}{2} + \ell_{t}^\mu\,, \notag \\
    & p_J^{\mu} = z Q \frac{n^{\mu}}{2} + \frac{p_T^2 }{z Q} \frac{\bar n^{\mu}}{2} + p_{t}^\mu\,, \quad  p_{\thickbar{J}}^{\mu} = (1-z) Q \frac{n^{\mu}}{2} + \frac{p_T^2 }{(1-z) Q} \frac{\bar n^{\mu}}{2} - p_{t}^\mu\,,
\end{align}
where we have introduced two light-like vectors, $n^\mu=(1,0,0,1)$ and $\bar n^\mu=(1,0,0,-1)$, and define $p_t^{\mu}$ such that $p_t^{\mu}{p_t}_{\mu} = - p_T^2$ with $\bm p_T = p_T (\cos \phi_J, \sin \phi_J)$. We denote transverse momenta relative to the photon-proton beam by the subscript $T$, while that relative to the jet direction is given the subscript $\perp$. 
Here, we assume $p_T^2\gg m_{\mathcal{Q}}^2$ and take $p_J^2= p_{\thickbar{J}}^2=0$. This allows us to derive the factorized cross section in the following section. Lastly, the parton-level Mandelstam variables can be defined as
\begin{align}
    & \hat{s} \equiv (p_g+q)^2 = (p_J+ p_{\thickbar{J}})^2 = \frac{p_T^2}{z(1-z)}\,,
    \\
    &\hat{t} \equiv (p_g-p_J)^2 = (q- p_{\thickbar{J}})^2 = - \frac{Q^2x\, z}{x_B}\,,
    \\
    &\hat{u} \equiv (p_g- p_{\thickbar{J}})^2 = (q-p_{J})^2 = -\frac{Q^2x\,(1-z)}{x_B}\,,
\end{align}
where $x$ is the momentum fraction of the proton carried by the gluon, and is given by
\begin{align}
    x=\frac{x_{B} D}{Q^{2} z(1-z)}, \quad \text { with } \quad D=Q^{2} z(1-z)+p_{T}^{2}\,.
\end{align}

\subsection{Factorization formula}

In the Breit frame, we define the dijet imbalance as $\bm q_T = \bm p_{JT} +  {\bm p}_{\thickbar{J} T}$. For this paper, we examine the back-to-back configuration where $q_T \ll  {p}_{\thickbar{J} T} \sim p_{JT} \equiv p_T$. Furthermore, we work in the kinematic regime where $m_{\mathcal{Q}} \lesssim p_T R \ll p_T$, with $R$ denoting the jet radius. Overall, in the region with the scale hierarchy as $q_T R \ll q_T \lesssim m_Q \lesssim p_T R \ll p_T$, the factorized expression for the proton-spin-independent cross section is given by
\begin{align}
    \label{eq:fact-kt-upol}
    \frac{d\sigma^{UU}}{dQ^2 d y d^2 \bm p_{T} d y_J d^2 \bm q_T} = & H(Q,y,p_T,y_J,\mu) \int d^2 \lambda_T\, d^2 k_T\, d^2 l_{\mathcal{Q}T}\, d^2 l_{\bar{\mathcal{Q}}T} S(\bm \lambda_T,\mu,\nu)  \\
    & \hspace{-1.5cm} \times \delta^{(2)}(\bm \lambda_T + \bm k_T + \bm{l}_{\mathcal{Q}T}+ \bm{l}_{\bar{\mathcal{Q}}T} - \bm q_T)\, f_{g/N}\left(x,k_T,\mu,\zeta/\nu^2\right) \,\notag \\
    & \hspace{-1.5cm} \times J_{\mathcal{Q}}(p_T R, m_{\mathcal{Q}}, \mu)\, S_{\mathcal{Q}}^{c}(\bm{l}_{\mathcal{Q}T}, R, m_{\mathcal{Q}}, \mu)\, J_{\bar{\mathcal{Q}}}(p_T R, m_{\mathcal{Q}}, \mu)\, S_{\bar{\mathcal{Q}}}^{c}(\bm{l}_{\bar{\mathcal{Q}}T}, R, m_{\mathcal{Q}}, \mu) \,.\notag 
\end{align}
Above, $y_J$ is the rapidity of the HF jet $J_{\mathcal{Q}}$ and is related to the kinematic variable $z$ through the relation $z=e^{y_J}p_T/Q$. In the factorization formula Eq. \eqref{eq:fact-kt-upol}, $S$ denotes the soft function while $f_{g/N}$ is the unpolarized gluon TMD PDF. Their perturbative one-loop expressions can be found in Sec.~\ref{sec:pdf-soft}. In the third line of Eq.~\eqref{eq:fact-kt-upol}, $J_{\mathcal{Q}}$ and $S_{\mathcal{Q}}^{c}$ are the massive quark jet and collinear-soft functions, which differ from the corresponding functions utilized in light jet production \cite{Buffing:2018ggv,Liu:2018trl,Liu:2020dct,Chien:2019gyf,Kang:2020xez}. In Secs.~\ref{sec:jetfunction} and \ref{sec:coftfunction}, we present their explicit calculations at next-to-leading order (NLO). The variables $\bm{k}_T$, $\bm{\lambda}_T$, and $\bm{l}_T$ label the transverse momenta associated with the collinear, soft, and collinear-soft modes. Finally, $\mu$ and $\nu$ are the factorization and rapidity scales, respectively, while $\zeta$ is the Collins-Soper parameter~\cite{Collins:2011zzd,Ebert:2019okf}. In the derivation of the above factorization formula we apply the narrow jet approximation with $R\ll1$. However, as shown in \cite{Jager:2004jh,Mukherjee:2012uz,Dasgupta:2016bnd,Liu:2018ktv} this approximation works well even for fat jets with radius $R\sim \mathcal{O}(1)$, and the power corrections of $\mathcal{O}(R^{2n})$ with $n>0$ can be obtained from the perturbative matching calculation. 

Fourier transforming to $b$-space, the factorized cross section becomes 
\begin{align}\label{eq:fac-unpl-rap}
    \frac{d\sigma^{UU}}{dQ^2 d y d^2 \bm p_{T} d y_J d^2 \bm q_T} = & H(Q,y,p_T,y_J,\mu)  \int \frac{d^2 b}{(2\pi)^2} e^{i\bm{b}\cdot \bm{q}_T} S(\bm b,\mu,\nu) \, f_{g/N}\left(x,b,\mu,\zeta/\nu^2\right) \,\notag \\
    & \hspace{-2.5cm} \times J_{\mathcal{Q}}(p_T R, m_{\mathcal{Q}}, \mu)\, S_{\mathcal{Q}}^{c}(\bm{b}, R, m_{\mathcal{Q}}, \mu)\, J_{\bar{\mathcal{Q}}}(p_T R, m_{\mathcal{Q}}, \mu)\, S_{\bar{\mathcal{Q}}}^{c}(\bm{b}, R, m_{\mathcal{Q}}, \mu) \,,
\end{align}
where soft function $S$ and the gluon TMD PDF $f_{g/N}$ both depend on the rapidity scale $\nu$. However, the soft function can be written as
\begin{align}\label{eq:TMD-PDF-redefin}
    S(\bm b,\mu,\nu) = \sqrt{S_{n\bar n}(b,\mu,\nu)} S(\bm b,\mu)\,,
\end{align}
where $S_{n\bar n}(b,\mu,\nu)$ is the soft function for Higgs production in $p+p$ collisions~\cite{Chiu:2012ir,Echevarria:2015uaa}, and the function $S(\bm b,\mu)$ on the right-hand side no longer depends on the rapidity scale $\nu$. Upon making this replacement, the factorized expression for the cross section can be written in terms of the properly-defined TMD gluon distribution~\cite{Collins:2011zzd} by noting that 
\begin{align}\label{e.prop_siv}
    f_{g/N}\left(x,b,\mu,\zeta/\nu^2\right) S(\bm b,\mu,\nu) = f_{g/N}^{\rm TMD}(x,b,\mu,\zeta) S(\bm b,\mu)\,.
\end{align}
Here, $f_{g/N}^{\rm TMD}(x,b,\mu,\zeta)$ on the right-hand side is defined as
\begin{align}
f_{g/N}^{\rm TMD}(x,b,\mu,\zeta) = f_{g/N}\left(x,b,\mu,\zeta/\nu^2\right) \sqrt{S_{n\bar n}(b,\mu,\nu)}\,.
\end{align}
This is the properly-defined gluon TMD PDF probed in Higgs production in $p+p$ collisions~\cite{Echevarria:2015uaa} and is thus the counterpart of the quark TMD PDF as probed in Drell-Yan lepton pair production. Finally, Eq.~\eqref{eq:fac-unpl-rap} can be expressed in the following form
\begin{align}
    \frac{d\sigma^{UU}}{dQ^2 d y d^2 \bm p_{T} d y_J d^2 \bm q_T} = & \,H(Q,y,p_T,y_J,\mu) \int \frac{d^2 b}{(2\pi)^2} e^{i\bm{b}\cdot \bm{q}_T} S(\bm b,\mu) \, f_{g/N}^{\rm TMD}(x,b,\mu,\zeta) \, \\
    & \hspace{-1.5cm} \times J_{\mathcal{Q}}(p_T R, m_{\mathcal{Q}}, \mu)\, S_{\mathcal{Q}}^{c}(\bm{b}, R, m_{\mathcal{Q}}, \mu)\, J_{\bar{\mathcal{Q}}}(p_T R, m_{\mathcal{Q}}, \mu)\, S_{\bar{\mathcal{Q}}}^{c}(\bm{b}, R, m_{\mathcal{Q}}, \mu) \,.\notag 
\end{align}
In the following sections we calculate the one-loop expressions for all the above functions. An important physical requirement is that the factorized cross section must be independent of the scale $\mu$--we verify this factorization-scale-independence in Sec. \ref{sec:rge}. 

Next, if one considers the scattering of an electron with a transversely-polarized proton with spin $\bm S_T$, Eq.~\eqref{eq:fact-kt-upol} can be generalized. In this case, the spin-dependent cross section is given by the sum
\begin{align}
    d\sigma(\bm S_T) = d\sigma^{UU} + d\sigma^{UT}(\bm S_T)\,,
\end{align}
where $d\sigma^{UT}$ depends on the gluon Sivers function. The full expressions for the leading twist gluon distributions are given in \cite{Mulders:2000sh}. Using these results, we can obtain the expression for the polarized cross section by simply replacing the unpolarized gluon TMD PDF in Eq.~\eqref{eq:fact-kt-upol} with the gluon Sivers function, namely
\begin{align}
    \label{eq:ftype}
    f_{g/N}\left(x,k_T,\mu,\zeta/\nu^2\right) \rightarrow \frac{1}{M}\epsilon_{\alpha\beta}\, S^{\alpha}_{T}\, k^{\beta}_{T}\, f_{1T,g/N}^{\perp, f}\left(x,k_T,\mu,\zeta/\nu^2\right).
\end{align}
Here, it is important to note that there exist both $f$- and $d$-type gluon Sivers functions, which are associated with different color configurations in the three-gluon correlator, i.e., involving the antisymmetric $f^{abc}$ and symmetric $d^{abc}$ structure constants of $SU(3)$, respectively. For details, see for instance \cite{Buffing:2013kca,Bomhof:2007xt}. In Eq.~\eqref{eq:ftype}, we have denoted the gluon Sivers function with the superscript $f$, which is used to indicate that it is $f$-type. We note that at LO, this process is only sensitive to the $f$-type function. Further details on this matter are provided in Sec.~\ref{Sec:Hard}. After making this substitution, the factorized cross section then reads
\begin{align}
    \label{eq:fact-kt-pol}
    \frac{d\sigma^{UT}(\bm S_T)}{dQ^2 d y d^2 \bm p_{T} d y_J d^2 \bm q_T} = & H^{\rm Sivers}(Q,y,p_T,y_J,\mu) \int d^2 \lambda_T\, d^2 k_T\, d^2 l_{\mathcal{Q}T}\, d^2 l_{\bar{\mathcal{Q}}T} S(\bm \lambda_T,\mu,\nu)  \\
    & \hspace{-1.5cm} \times \delta^{(2)}(\bm \lambda_T + \bm k_T + \bm{l}_{\mathcal{Q}T}+ \bm{l}_{\bar{\mathcal{Q}}T} - \bm q_T)\, \frac{1}{M}\epsilon_{\alpha\beta}\, S^{\alpha}_{T}\, k^{\beta}_{T}\, f_{1T,g/N}^{\perp, f}\left(x,k_T,\mu,\zeta/\nu^2\right) \,\notag \\
    & \hspace{-1.5cm} \times J_{\mathcal{Q}}(p_T R, m_{\mathcal{Q}}, \mu)\, S_{\mathcal{Q}}^{c}(\bm{l}_{\mathcal{Q}T}, R, m_{\mathcal{Q}}, \mu)\, J_{\bar{\mathcal{Q}}}(p_T R, m_{\mathcal{Q}}, \mu)\, S_{\bar{\mathcal{Q}}}^{c}(\bm{l}_{\bar{\mathcal{Q}}T}, R, m_{\mathcal{Q}}, \mu) \,,\notag 
\end{align}
where $H^{\rm Sivers}$ denotes the hard function for the polarized process, and this expression can once again be written as a Fourier transform by defining
\begin{align}
     \frac{ib^\beta}{2}f_{1T,g/N}^{\perp, f}(x,b,\mu,\zeta/\nu^2) = \int d^2 k_T e^{-i \bm{b}\cdot \bm{k}_T}\frac{k^{\beta}_{T}}{M} f_{1T,g/N}^{\perp, f}(x,k_T,\mu,\zeta/\nu^2)\,.
\end{align}
Finally, the factorization formula for the polarized differential cross section becomes
\begin{align}
    \frac{d\sigma^{UT}(\bm S_T)}{dQ^2 d y d^2 \bm p_{T} d y_J d^2 \bm q_T} = & H^{\textrm{Sivers}}(Q,y,p_T,y_J,\mu)  \int \frac{d^2 b}{(2\pi)^2} e^{i\bm{b}\cdot \bm{q}_T} S(\bm b,\mu) \, \\
    & \hspace{-1.5cm} \times \frac{i}{2}(\epsilon_{\alpha\beta}\,S_T^\alpha\, b^\beta)f_{1T,g/N}^{\perp, f}(x,b,\mu,\zeta) \,\notag \\
    & \hspace{-1.5cm} \times J_{\mathcal{Q}}(p_T R, m_{\mathcal{Q}}, \mu)\, S_{\mathcal{Q}}^{c}(\bm{b}, R, m_{\mathcal{Q}}, \mu)\, J_{\bar{\mathcal{Q}}}(p_T R, m_{\mathcal{Q}}, \mu)\, S_{\bar{\mathcal{Q}}}^{c}(\bm{b}, R, m_{\mathcal{Q}}, \mu) \nn \,.
\end{align}
Here, we have applied the redefinition Eq.~\eqref{eq:TMD-PDF-redefin} to obtain the rapidity-scale-independent gluon Sivers function
\begin{align}\label{e.sivers}
f_{1T,g/N}^{\perp, f}(x,b,\mu,\zeta) \equiv f_{1T,g/N}^{\perp, f}(x,b,\mu,\zeta/\nu^2) \sqrt{S_{n\bar n}(b,\mu,\nu)}\,.
\end{align}

\subsection{Hard function}\label{Sec:Hard}
In the unpolarized process, the LO hard function is determined by the tree-level cross section for dijet production in DIS, which is expressed as \cite{Korner:1988bp,Mirkes:1997ru}
\begin{align}
    H(Q,y,p_T,y_J,\mu) = & \frac{\alpha_{\rm em}^2\alpha_s Q_f^2C_F C_A}{4\pi Q^2 y^2 S_{\ell P}} \Big\{ \left[1+(1-y)^2\right]H^{U,U+L} - y^2\, H^{U,L}  \\
     & \hspace{4.7cm}- (2-y) \sqrt{1-y}\, H^{U,I}  + 2(1-y)\, H^{U,T}\Big\}\,, \notag
\end{align}
where $\alpha_{\rm em}$ is the fine structure constant and $Q_f$ denotes the fractional charge of the HF quark. On the right-hand side, the first superscript $U$ indicates that the incoming gluon is unpolarized, while the second superscripts $\{U+L,L,I,T\}$ correspond to the different helicity states of the off-shell photon. Explicitly, the functions $H^{U,i}$ are expressed as
\begin{align}
& H^{U,I} = \cos(\phi_J) H_{\cos(\phi_J)}^{U,I},\quad
H^{U,T} = \cos(2\phi_J) H_{\cos(2\phi_J)}^{U,T}, \notag\\
& H_{\cos(\phi_J)}^{U,I} =\frac{2z(1-2z)p_{T} Q}{D^2} \left( -1 + \frac{2p_{T}^2}{D}\right), \quad
   H_{\cos(2\phi_J)}^{U,T} = \frac{2z\,p_{T}^2}{D^2}\left(1-\frac{p_{T}^2}{D} \right), \notag \\
& H^{U,U+L} = \frac{1}{D}\left(\frac{1}{2-2z}-z\right) + \frac{2zp_{T}^2}{D^2}  - \frac{z p_{T}^2}{D^3}\left(2p_{T}^2 + Q^2(1-2z)^2\right), \notag\\ 
& H^{U,L} = \frac{4 z p_{T}^2}{D^2}\left( 1 -\frac{ p_{T}^2}{D}\right)\,.
\end{align}
One immediately sees that the functions $H^{U,I}$ and $H^{U,T}$ vanish upon integrating out the azimuthal angle $\phi_J$ of the jet. As such, these contributions do not play a role in our numerical calculations.  

The expression for the hard anomalous dimension can be obtained from the calculation of the $3$-jet process $\gamma^*\to q \bar q g$ at lepton collisions \cite{Giele:1991vf,Arnold:1988dp}. The hard anomalous dimension can also be read from the general structures in \cite{Catani:1998bh,Becher:2009cu} and is given as
\begin{align}\label{eq:hard-ad}
    \Gamma^h(\alpha_s) &= C_A \gamma^{\rm cusp}(\alpha_s) \ln\left( \frac{\hat{u} \, \hat{t}}{\hat{s} \, \mu^2}\right) - 2 C_F \gamma^{\rm cusp}(\alpha_s) \ln\left( \frac{\mu^2}{\hat{s}}\right)+4\gamma^q(\alpha_s) +2 \gamma^g(\alpha_s)\,, 
\end{align}
where $\gamma^{\rm cusp}$ is the cusp anomalous dimension, while $\gamma^q$ and $\gamma^g$ represent the single logarithmic anomalous dimensions for the quark and gluon, respectively. With this anomalous dimension, one can then perform resummation by solving the following renormalization group (RG) equation for the hard function  
\begin{align}
    \frac{d}{d\ln\mu}\ln H(\mu) = \Gamma^h(\alpha_s)\,,
\end{align}
where, for brevity, we maintain only the scale $\mu$-dependence in the hard function. We note that in order to perform the evolution at next-to-leading logarithmic (NLL) accuracy, the cusp anomalous dimension is needed at two-loop order and the single logarithmic anomalous dimensions are needed at one-loop order. The values of these expressions are
\begin{align}\label{eq:one-loop-hard-ad}
    &\gamma^{\rm cusp}_0=4\,, \quad \gamma_1^{\rm cusp} = \left(\frac{268}{9} - \frac{4\pi^2}{3}\right)C_A - \frac{40}{9}C_F n_f\,,\notag  \\
    & \gamma_0^q=-3C_F\,,\quad \gamma_0^g=-\beta_0,\quad \beta_0 = \frac{11}{3}C_A - \frac{4}{3}T_Fn_f\,,
\end{align}
where we have organized the perturbative expansion of each anomalous dimension as 
\begin{align}
\gamma(\alpha_s) = \frac{\alpha_s}{4\pi}\gamma_0+\left(\frac{\alpha_s}{4\pi}\right)^2\gamma_1 + \mathcal{O}(\alpha_s^3)\,.
\end{align}

\begin{figure}
    \begin{center}
    \includegraphics[height = 1.in,valign = c]{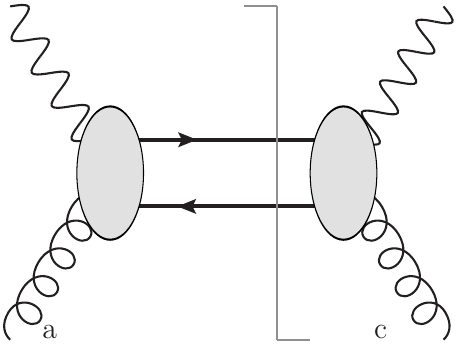}
    $ = C_u\, \delta^{ac}\, h(Q,y,p_T,y_J,\mu)$
    \\
    \vspace{0.5cm}
    \includegraphics[height = 1.in,valign = c]{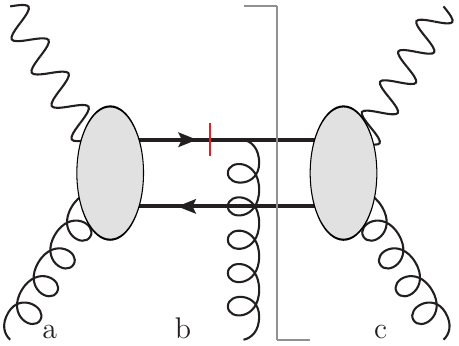} + 
    \includegraphics[height = 1.in,valign = c]{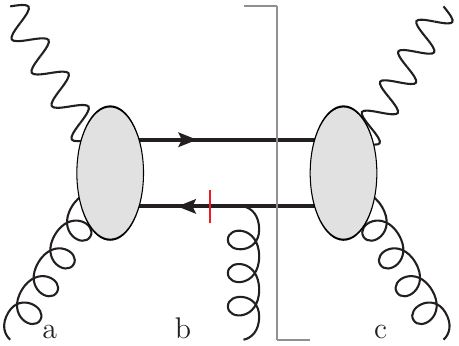}
    $ = \left(C_1+C_2\right) \left(-i f^{abc}\right) h(Q,y,p_T,y_J,\mu)$
    \end{center}
    \caption{Top: Unpolarized hard Feynman diagram for HF dijet production. Bottom: Polarized hard diagram for HF dijet production. The red lines in the polarized case indicate the location of a soft pole. }
    \label{fig:Hard-Pol}
\end{figure}

For the polarized process, we must consider the process-dependence of the corresponding gluon Sivers functions~\cite{Buffing:2013kca}. Such process-dependence can be computed via the attachment of an additional gluon originating from the gauge link in the definition of the Sivers function. This additional gluon is responsible for the soft pole that generates Sivers asymmetry. This method is widely used in computing the process-dependence of the quark Sivers function, see e.g.~\cite{Qiu:2007ey}, which gives the same results as shown in~\cite{Buffing:2013kca,Bomhof:2006ra}. In Fig.~\ref{fig:Hard-Pol}, the soft poles are represented by red lines. We note that for both the polarized and unpolarized cases, the hard functions can be expressed as matrices in color space \cite{Kang:2020xez}. For more complicated processes, the relationship between the polarized and unpolarized hard matrices is non-trivial. However, for the $\gamma^*q\rightarrow \mathcal{Q}\bar{\mathcal{Q}}$ process, the color space is one-dimensional and, therefore, the polarized hard function can be simply written as
\begin{align}
    \label{eq:H-Sivers}
    H^{\rm{Sivers}}(Q,y,p_T,y_J,\mu) = \left(C_1+C_2\right)\, h(Q,y,p_T,y_J,\mu)\,.
\end{align}
Here, $C_1$ and $C_2$ are the color factors for the polarized hard process associated with the attachment of the additional gluon to the HF quark and anti-quark~\cite{Yuan:2008vn,Kang:2008qh,DAlesio:2020eqo}, respectively. The function $h(Q,y,p_T,y_J,\mu)$ is the kinematic part of the hard function. For the unpolarized case, the hard function can be written as
\begin{align}
    \label{eq:H-Upol}
    H(Q,y,p_T,y_J,\mu) = C_u\, h(Q,y,p_T,y_J,\mu)\,,
\end{align}
where factor $C_u$ is the color factor associated with the unpolarized hard process. We find that at LO for this process, the attachment of this additional gluon originating from the gauge link in the definition of the Sivers function produces color configurations which are proportional to $(-if^{abc})$. This analysis indicates that while there are both $d$- and $f$-type gluon Sivers functions \cite{Buffing:2013kca}, this process at LO is only sensitive to the $f$-type gluon Sivers function. In addition to this, 
while the term $(-if^{abc})$ in Fig.~\ref{fig:Hard-Pol} appears in the hard function, 
this term should be absorbed into the definition of the gluon Sivers function as it originates from the Wilson line in the adjoint representation. Similarly, the term $\delta^{ac}$ in that figure should be absorbed into the definition of the unpolarized TMD PDF. For this process, we find that $\left(C_1+C_2\right) = C_u$. As a result, the polarized and unpolarized hard functions are equal and the hard anomalous dimension is unchanged for the polarized case.

\subsection{TMD PDFs and the global soft function}\label{sec:pdf-soft}
In order to regulate the rapidity divergences in the TMD PDFs and the global soft function, we use the $\eta$ regulator of \cite{Chiu:2012ir}. The expression for the unsubtracted gluon TMD in this regularization scheme can be obtained from \cite{Kang:2017glf}, and is given by 
\begin{align}
    \label{eq.TMDPDFgg}
    & f_{g/g}^{\rm NLO}(x,b,\mu,\zeta/\nu^2) =  \delta(1-x) \\
    & \hspace{2.8cm} +\frac{\alpha_s}{4\pi} C_A\left[\frac{4}{\eta}\left(\frac{1}{\epsilon}+\ln\frac{\mu^2}{\mu_b^2}\right) +\frac{2}{\epsilon}\left(\ln\frac{\nu^2}{\left(n\cdot p_g\right)^2}+\frac{\beta_0}{ C_A}\right) \right] \delta(1-x) \nn \\
    & \hspace{2.8cm}-\frac{\alpha_s}{4\pi}\left[ \frac{2}{\epsilon}+2\ln\frac{\mu^2}{\mu_b^2} \right]P_{gg}(x) \nn \\
    & \hspace{2.8cm}+\frac{\alpha_s}{4\pi}C_A\left[ \ln \frac{\mu^2}{\mu_b^2}  \left( \ln \frac{\nu^2}{\left(n\cdot p_g\right)^2}  +\frac{\beta_0}{ C_A}\right)\delta(1-x)\right]\,,\nn
    \\
    \label{eq.TMDPDFgq}
    & f_{g/q}^{\rm NLO}(x, b, \mu,\zeta/\nu^2) = -\frac{\alpha_s}{4\pi}\left[ \frac{2}{\epsilon}+2\ln \frac{\mu^2}{\mu_b^2}  \right]P_{gq}(x)+\frac{\alpha_s}{4\pi} C_F \left(2x\right)\,,
\end{align}
where we have set $\zeta = \left(n\cdot p_g\right)^2$~\cite{Ebert:2019okf}, the scale $\mu_b$ is defined as $\mu_b = 2e^{-\gamma_E}/b$, with $b$ being the magnitude of the two-dimensional vector transverse to the beam direction $\bm{b}=b\left(\cos\phi_b ,\sin \phi_b\right)$, and
\begin{align}
    & P_{gg}(x) = 2 C_A\left[\frac{x}{(1-x)}_+ +\frac{1-x}{x}+x(1-x)\right]+\frac{\beta_0}{2}\delta(1-x)\,,
    \\
    & P_{gq}(x) = C_F\frac{1+(1-x)^2}{x}\,,
\end{align}
are the collinear splitting kernels. The term in the third line of Eq.~\eqref{eq.TMDPDFgg} and the analogous term in Eq.~\eqref{eq.TMDPDFgq} contain the infrared divergences which are to be matched to the collinear PDF. The RG equations for the gluon TMD PDF are then 
\bea
\frac{d}{d\ln\mu}\ln f_{g/N}(x,b,\mu,\zeta/\nu^2) =\, & \Gamma^{f_g}(\alpha_s)\,,
\\
\frac{d}{d\ln\nu}\ln f_{g/N}(x,b,\mu,\zeta/\nu^2) =\, & \gamma_{\nu}^{f_g}(\alpha_s)\,.
\eea
Here the anomalous dimensions are given by
\begin{align}
    \label{eq:beam-ad}
    \Gamma^{f_g}(\alpha_s) =  C_A \gamma^{\textrm{cusp}}(\alpha_s) \ln\frac{\nu^2}{(n\cdot p_g)^2} -2\gamma^{f_g}(\alpha_s)\,,\quad
    \gamma_{\nu}^{f_g}(\alpha_s) = \frac{\alpha_s}{\pi}C_A \ln\frac{\mu^2}{\mu_b^2}+\mathcal{O}(\alpha_s^2)\,,
\end{align}
with $\gamma^{f_g}_0 = \gamma_0^g$, which is given in Eq.~\eqref{eq:one-loop-hard-ad}.

On the other hand, we define the soft function as
\begin{align}
\label{eq:gsoft}
S^{\rm LO }(b,\mu,\nu) &= 1\,,\\
S^{\rm{NLO} }(\bm b,\mu,\nu) &= -\frac{C_A}{2}\,\mathcal{I}_{BJ} -\frac{C_A}{2}\,\mathcal{I}_{B\thickbar{J}} + \left(\frac{C_A}{2}-C_F\right)\,\mathcal{I}_{J\thickbar{J}}\,,
\label{eq:gsoft1loop}
\end{align}
where the soft integrals $\mathcal{I}_{ij}$ are given by
\begin{align}
    \mathcal{I}_{ij}
    = \frac{\alpha_s \mu^{2\epsilon} \pi^\epsilon e^{\epsilon\gamma_E}}{\pi^2} \int d^dk\, \delta^+(k^2)e^{- i \,k\cdot b} \frac{n_i\cdot n_j}{(n_i\cdot k)(n_j\cdot k)} \frac{\nu^\eta}{|k^+-k^-|^\eta} \,.
\end{align}
Here, $i$ and $j$ are either $B$, $J$, or $\thickbar{J}$, where $B$ denotes the beam direction while $J$ and $\thickbar{J}$ denote the directions of the jet initiated by the HF quark and anti-quark, respectively.
The expressions for the $\phi_b$-dependent beam-jet soft function integrals are given in \cite{Buffing:2018ggv} as
\begin{align}
\label{eq:soft-int1}
\mathcal{I}_{BJ} = &\, \frac{\alpha_s}{4\pi}\Bigg{[} \frac{4}{\eta}\left( \frac{1}{\epsilon}+\ln\frac{\mu^2}{\mu_b^2} \right) -\frac{4}{\epsilon^2}-\frac{2}{\epsilon}\left( -2 y_J-\ln\frac{\nu^2}{\mu^2} + 2 \ln \frac{2i \mu\,c_{bJ}}{\mu_b} \right) \Bigg{]}  + \mathcal{I}_{BJ}^{ \textrm{fin}},
\\
\label{eq:soft-int2}
\mathcal{I}_{B\thickbar{J}} = &\, \frac{\alpha_s}{4\pi}\Bigg{[} \frac{4}{\eta}\left( \frac{1}{\epsilon}+\ln\frac{\mu^2}{\mu_b^2} \right) -\frac{4}{\epsilon^2}-\frac{2}{\epsilon}\left( -2 y_{\thickbar{J}}-\ln\frac{\nu^2}{\mu^2} +2 \ln \frac{-2i \mu\,c_{bJ}}{\mu_b} \right) \Bigg{]}  + \mathcal{I}_{B\thickbar{J}}^{ \textrm{fin}},
\end{align}
with $c_{bJ}=\cos(\phi_b-\phi_J)$ and where the rapidity of jets are
\begin{align}
    y_J = \ln\frac{z Q}{p_T},
    \qquad
    y_{\thickbar{J}} = \ln\frac{(1-z)Q}{p_T}\,.
\end{align}
Here, the terms marked by ``$\textrm{fin}$" in their superscripts denote the finite contributions of their respective functions. While the divergent pieces are required for the purposes of resummation, the finite pieces are only needed at NLO. Since we perform our analysis at NLL accuracy, these terms are not needed for this study. Recently in \cite{delCastillo:2020omr}, the authors derive the following the jet-jet soft function integral which contains no rapidity divergence. 
This integral can be written as 
\begin{align}
    \label{eq:soft-int3}
    \mathcal{I}_{J\thickbar{J}} = \frac{\alpha_s}{4\pi}\left[ -\frac{4}{\epsilon^2} +\frac{4}{\epsilon}\left(\ln \frac{\hat s}{p_T^2}-\ln\frac{\mu^2}{\mu_b^2} -\ln\left( 4c_{bJ}^2\right) \right)\right]  + \mathcal{I}_{J\thickbar{J}}^{ \textrm{fin}}\,.
\end{align}
From Eqs.~\eqref{eq:soft-int1}, \eqref{eq:soft-int2}, and \eqref{eq:soft-int3}, we obtain the following expressions for the soft anomalous dimensions
\begin{align}
    \label{eq:soft-ad}
    \Gamma^s(\alpha_s) & =  2 C_F\,\gamma^{\rm{cusp}}(\alpha_s)\ln \frac{\mu^2}{\mu_b^2}  -C_A\,\gamma^{\rm{cusp}}(\alpha_s) \ln \frac{\nu^2}{\mu^2} + \gamma^s(\alpha_s)\,, \\
    \label{eq:soft-rap-ad}
    \gamma_\nu^s(\alpha_s) &= -\frac{\alpha_s}{\pi}C_A \ln\frac{\mu^2}{\mu_b^2}+\mathcal{O}(\alpha_s^2)\,,
\end{align}
with the one-loop single logarithmic anomalous dimension
\begin{align}
    \gamma_0^s = 8 C_F \ln\left( 4 c_{bJ}^2 \right) +4(C_A-2 C_F)\ln\frac{\hat{s}}{p_T^2}+4\,C_A \ln\frac{\hat{s}}{Q^2}\,.
\end{align}
As expected, we see that the one-loop rapidity anomalous dimensions of the TMD PDF and the soft function fulfill the condition
\begin{align}
    \gamma_{\nu,0}^{f_g}+\gamma_{\nu,0}^s = 0\,.
\end{align}
Thus, the product of $f_{g/N}(x,b,\mu,\zeta/\nu^2)$ and $S(\bm b,\mu,\nu)$ is $\nu$-independent, and we can construct the properly-defined gluon TMD PDF as in Eq.~\eqref{e.prop_siv}. 

Here we find that the soft function depends not only the magnitude but also the direction of the vector $\bm b$.  As shown in Sec. \ref{sec:coftfunction} a similar structure also shows up in the collinear-soft function, and the $\phi_b$ dependence in the anomalous dimensions will cancel out between these two functions.  However, after taking into account the evolution between the soft and collinear-soft function, one finds that the $\phi_b$ integral is divergent in some phase space region.  In order to avoid such divergences we apply methods in \cite{Buffing:2018ggv,Kang:2020xez} where one first performs an averaging over the $\phi_b$ angle in both soft and collinear-soft function. We note that this method does not change the RG invariance as shown in Eqs. \eqref{eq:rge1} and \eqref{eq:rge2}. In addition, as discussed in \cite{Chien:2019gyf} no significant numerical effects between different methods are observed in the NLL resummation calculation.


The $\phi_b$-averaged soft function can be constructed from Eqs.~\eqref{eq:gsoft} and \eqref{eq:gsoft1loop} by replacing the soft integrals with 
\begin{align}
\bar{\mathcal{I}}_{BJ}
= \frac{\alpha_s}{4\pi}&\left[2\left(\frac{2}{\eta}+\ln\frac{\nu^2}{\mu^2}+2y_J\right)\left(\frac{1}{\epsilon}+\ln\frac{\mu^2}{\mu_b^2}\right) -\frac{4}{\epsilon^2} -\frac{2}{\epsilon}\ln\frac{\mu^2}{\mu_b^2}+\frac{\pi^2}{3}\right]\,,\\
\bar{\mathcal{I}}_{B\thickbar{J}}
= \frac{\alpha_s}{4\pi}&\left[2\left(\frac{2}{\eta}+\ln\frac{\nu^2}{\mu^2}+2y_{\thickbar{J}}\right)\left(\frac{1}{\epsilon}+\ln\frac{\mu^2}{\mu_b^2}\right) -\frac{4}{\epsilon^2} -\frac{2}{\epsilon}\ln\frac{\mu^2}{\mu_b^2}+\frac{\pi^2}{3}\right]\,,\\
\bar{\mathcal{I}}_{J\thickbar{J}}
=\frac{\alpha_s}{2\pi}&\,\bigg[4\left(\frac{1}{\epsilon}+\ln\frac{\mu^2}{\mu_b^2}\right)\ln\big(2\cosh(\Delta y/2)\big) -\frac{2}{\epsilon^2}-\frac{2}{\epsilon}\ln\frac{\mu^2}{\mu_b^2} -\ln^2\frac{\mu^2}{\mu_b^2} + \Delta y^2  \notag\\
&- 4\ln^2\big(2\cosh(\Delta y/2)\big)+\frac{\pi^2}{6}\bigg]\,,
\label{eq:gsoftIexp-avg}
\end{align}
where we have placed a bar over these integrals in order to distinguish them from the $\phi_b$-dependent ones and have defined $\Delta y=y_J-y_{\thickbar{J}}$. These results are the same as the soft function calculated in \cite{Hornig:2017pud}. Therefore, the anomalous dimensions for the averaged case are
\begin{align}
    \label{eq:soft-ad-avg}
    \bar{\Gamma}^s(\alpha_s) & =  2 C_F\,\gamma^{\rm{cusp}}(\alpha_s)\ln \frac{\mu^2}{\mu_b^2}  -C_A\,\gamma^{\rm{cusp}}(\alpha_s) \ln \frac{\nu^2}{\mu^2} + \bar{\gamma}^s(\alpha_s)\,, \\
    \label{eq:soft-rap-ad-avg}
    \bar{\gamma}_\nu^s(\alpha_s) & = -\frac{\alpha_s}{\pi}C_A \ln\frac{\mu^2}{\mu_b^2} + \mathcal{O}(\alpha_s^2)\,,
\end{align}
where the one-loop single logarithmic anomalous dimensions is
\begin{align}
    \bar \gamma_0^s = 4(C_A-2 C_F)\ln\frac{\hat{s}}{p_T^2}+4C_A \ln\frac{\hat{s}}{Q^2}\,.
\end{align}
By comparing Eqs.~\eqref{eq:soft-rap-ad} and \eqref{eq:soft-rap-ad-avg}, one can see the rapidity anomalous dimension is unchanged. Therefore in the $\phi_b$-averaged case, we can once again write the factorized expression in terms of the properly defined TMD PDFs.

\subsection{Massive quark jet function}\label{sec:jetfunction}

In this section, we discuss the calculation of the massive quark jet function at NLO. The massive quark jet function has been investigated in detail for various observables. For example, the factorization formula for the massive event shape distribution involves such a jet function, as the jet and heavy quark masses are of similar magnitude \cite{Fleming:2007qr,Fleming:2007xt,Lepenik:2019jjk,Bris:2020uyb}. The corresponding jet function has been calculated to two-loop order \cite{Hoang:2019fze}. Furthermore, the semi-inclusive massive quark jet fragmentation function has been calculated at NLO and applied to inclusive jet production \cite{Dai:2018ywt,Li:2018xuv}. Recently, the one-loop expression for the so-called unmeasured massive quark jet function has been presented in \cite{Kim:2020dgu}. 

\begin{figure}[t]
    \centering
    \begin{overpic}[width=0.85\linewidth]{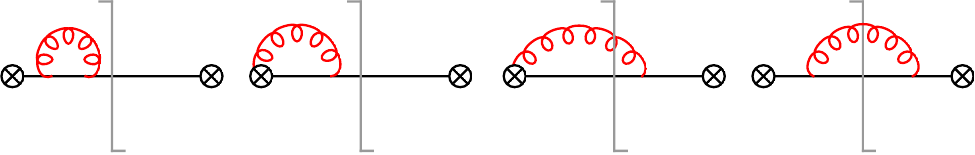}
    \color{black}\put(310,20){$q$}
    \end{overpic}
    \vspace{0.2cm}
    \caption{Sample Feynman diagrams contributing to the massive quark jet function $J_{\mathcal{Q}}$ at one-loop order in perturbation theory. The virtual corrections $J_{\mathcal{Q}}^{\rm NLO,V}$ are displayed in the first two diagrams, where each contain only a single cut propagator. The remaining diagrams involving two cut propagators represent the real corrections $J_{\mathcal{Q}}^{\rm NLO,R}$.}
    \label{fig:jetfunction}
\end{figure}

The global jet anomalous dimension can be obtained from the divergent terms of the unmeasured massive quark jet function. As shown in Fig.~\ref{fig:jetfunction}, the one-loop calculation involves two types of diagrams: $J_{\mathcal{Q}}^{\rm NLO,V}$ and $J_{\mathcal{Q}}^{\rm NLO,R}$, where $J_{\mathcal{Q}}^{\rm NLO,V}$ contains only single cut propagators and is thereby unconstrained by the jet algorithm. Explicitly, it is written as
\begin{align}\label{eq:jet-virtual}
    J_{\mathcal{Q}}^{\rm NLO,V} = \frac{\alpha_s}{4\pi} C_F \left[ \frac{2}{\epsilon^2} + \frac{1}{\epsilon}\left(1+ 2\ln \frac{\mu^2}{m_{\mathcal{Q}}^2}\right)+\left(1+\ln \frac{\mu^2}{m_{\mathcal{Q}}^2}\right)\ln \frac{\mu^2}{m_{\mathcal{Q}}^2}+4+\frac{\pi^2}{6} \right]\,,
\end{align}
where the heavy quark mass $m_{\mathcal{Q}}$ is the only physical scale involved. Since the real contribution $J_{\mathcal{Q}}^{\rm NLO,R}$ is constrained by the jet algorithm, it will depend on the jet scale $p_TR$ in addition to $m_{\mathcal{Q}}$. In this work, we define the HF quark four-momentum $q^\mu$ with $q^2=m_{\mathcal{Q}}^2$, which is known as the M-scheme \cite{Bris:2020uyb}.
We note that in the hierarchy of scales we are considering, the constraint of the anti-$k_T$ algorithm \cite{Cacciari:2008gp} is independent of the HF quark mass $m_{\mathcal{Q}}$ and is in fact identical to that for massless partons \cite{Dai:2018ywt}, namely
\begin{align}\label{eq:antikt}
    \Theta_{\operatorname{anti-}\!k_T}=\theta\left[\left(\frac{q^-\left(\omega_J-q^-\right)}{\omega_J}\right)^2\left(\frac{R}{2\,\mathrm{cosh}\,y_J}\right)^2-\bm{q}_{\perp}^2\right]\,,
\end{align}
where $q^\mu=(q^+,q^-,\bm{q}_\perp)$ is the four-momentum of the HF quark and $\omega_J$ is the large component of the jet four-momentum. The jet scale $p_TR$ emerges in Eq.~\eqref{eq:antikt} upon noting $\omega_J=2\,p_T\,\mathrm{cosh}\,y_J$. In the phase space integral, we expand the integrated momentum $q$ along the jet direction with $q^+ = (m_{\mathcal{Q}}^2+\bm q_\perp^2)/q^-$ given by the power counting requirement $p_T R\sim m_{\mathcal{Q}}$. Explicitly, we have 
\begin{align}\label{eq:jet-full}
    J_{\mathcal{Q}}^{\rm NLO,R}(p_T R,m_{\mathcal{Q}},&\epsilon) =  \frac{\alpha_s C_F e^{\epsilon \gamma_E}\mu^{2\epsilon}}{2\pi \Gamma(1-\epsilon)} \int \frac{d q^-}{\omega_J} \frac{d \bm q_\perp^2}{\bm q_\perp^{2\epsilon}} \Bigg[ \frac{q^-}{\omega_J-q^-}  \frac{2\bm q_\perp^2 \omega_J^4}{[\bm q_\perp^2 \omega_J^2 + m_{\mathcal{Q}}^2 (\omega_J-q^-)^2]^2} \notag \\
    & + (1-\epsilon) \frac{\omega_J(\omega_J-q^-)}{\bm q_\perp^2 \omega_J^2 + m_{\mathcal{Q}}^2 (\omega_J-q^-)^2} \Bigg] \theta( \omega_J - q^-) \Theta_{\operatorname{anti-}\!k_T}\notag \\ 
    &= \frac{\alpha_s}{4\pi}C_F \left[ - 2 \ln\left( \frac{m_{\mathcal{Q}}^2+p_T^2 R^2}{m_{\mathcal{Q}}^2}\right) + 2 - \frac{2m_{\mathcal{Q}}^2}{m_{\mathcal{Q}}^2+ p_T^2 R^2}  \right]\frac{1}{\epsilon} + J_{\mathcal{Q}}^{\mathrm{R,fin}}\,,
\end{align}
where only a single divergence is exhibited, as the heavy quark mass $m_{\mathcal{Q}}$ acts as a regulator of the overlapping soft and collinear regions of phase space. After combining the real and virtual contributions, the logarithmic dependence on the quark mass $m_{\mathcal{Q}}$ cancels out. 

The one-loop global jet renormalization constant then reads
\begin{align}\label{eq:jet-ad}
    Z^{J_{\mathcal{Q}}} = 1 + \frac{\alpha_s}{4\pi} C_F \left[ \frac{2}{\epsilon^2} + \frac{1}{\epsilon}\left( 2 \ln \frac{\mu^2}{m_{\mathcal{Q}}^2 + p_T^2R^2} + 3 - \frac{2m_{\mathcal{Q}}^2}{m_{\mathcal{Q}}^2 + p_T^2 R^2} \right) \right]\,,
\end{align}
where, again, we observe that the heavy quark mass $m_{\mathcal{Q}}$ only affects the single pole structure. We further note that as $m_{\mathcal{Q}} \to 0$, the massive quark jet renormalization constant reduces to that of the massless jet, $Z^{J_{\mathcal{Q}}} \to Z^{J_q}$.  This gives us the following expression for the global jet anomalous dimension
\begin{align}\label{eq:jf-ad}
    \Gamma^{j_{\mathcal{Q}}}(\alpha_s) = - C_F \gamma^{\rm cusp}(\alpha_s) \ln \frac{m_{\mathcal{Q}}^2+p_T^2R^2}{\mu^2} + \gamma^{j_{\mathcal{Q}}}(\alpha_s)\,,
\end{align}
with the one-loop single logarithmic anomalous dimension as
\begin{align}
    \gamma_0^{j_{\mathcal{Q}}} = 2 C_F \left( 3 - \frac{2m_{\mathcal{Q}}^2}{m_{\mathcal{Q}}^2+ p_T^2R^2} \right)\,,
\end{align}
where the first term in the brackets is shared by the massless quark jet function and the second term constitutes the finite quark mass correction. 
Finally, the renormalized HF jet function is given by the following 
\begin{align}\label{eq:massive-jet}
    J_{\mathcal{Q},{\rm NLO}}^{\mathrm{ren.}}(p_TR,m_{\mathcal{Q}},\mu)=&\,\frac{\alpha_s}{4\pi}C_F\Bigg[\left(3-\frac{2m_{\mathcal{Q}}^2}{m_{\mathcal{Q}}^2+p_T^2R^2}\right)\ln \frac{\mu^2}{m_{\mathcal{Q}}^2+p_T^2R^2}  \\
    &\hspace{2.5cm}+\ln^2\frac{\mu^2}{m_{\mathcal{Q}}^2+p_T^2R^2}+13-\frac{3\pi^2}{2}+\mathcal{F}(p_TR,m_{\mathcal{Q}})\Bigg]\,,\notag
\end{align}
where the function $\mathcal{F}(p_TR,m_{\mathcal{Q}})$ can be expressed as
\begin{align}
    \mathcal{F}(p_TR,m_{\mathcal{Q}})=&\,\pi^2-4\,\mathrm{Li}_2\left(-\frac{m_{\mathcal{Q}}^2}{p_T^2R^2}\right)+2\left(1-\ln\frac{m_{\mathcal{Q}}^2}{p_T^2R^2}\right)\ln\frac{m_{\mathcal{Q}}^2+p_T^2R^2}{p_T^2R^2} \notag \\
    &- \frac{2m_{\mathcal{Q}}^2}{m_{\mathcal{Q}}^2+p_T^2R^2}\ln\frac{m_{\mathcal{Q}}^2}{p_T^2R^2}-\frac{m_{\mathcal{Q}}^2}{p_T^2R^2}\ln\frac{m_{\mathcal{Q}}^2+p_T^2R^2}{m_{\mathcal{Q}}^2} \notag \\
    &-4\bigg[\frac{m_{\mathcal{Q}}}{p_TR}\left(1+\frac{m_{\mathcal{Q}}^2}{m_{\mathcal{Q}}^2+p_T^2R^2}\right)+\mathrm{Cot}^{-1}\left(\frac{m_{\mathcal{Q}}}{p_TR}\right)\bigg]\mathrm{Cot}^{-1}\left(\frac{m_{\mathcal{Q}}}{p_TR}\right)\,.
\end{align}
This expression for the HF jet function is equivalent to the semi-analytic form presented in \cite{Kim:2020dgu}, and one can see that as $m_{\mathcal{Q}}\rightarrow0$, we have $\mathcal{F}\rightarrow0$ and, therefore, $J_{\mathcal{Q}}\rightarrow J_q$. Hence, the massive quark jet function behaves as expected in the massless limit.

\subsection{Collinear-soft function}\label{sec:coftfunction}

\begin{figure}[t]
    \centering
    \begin{overpic}[width=0.75\linewidth]{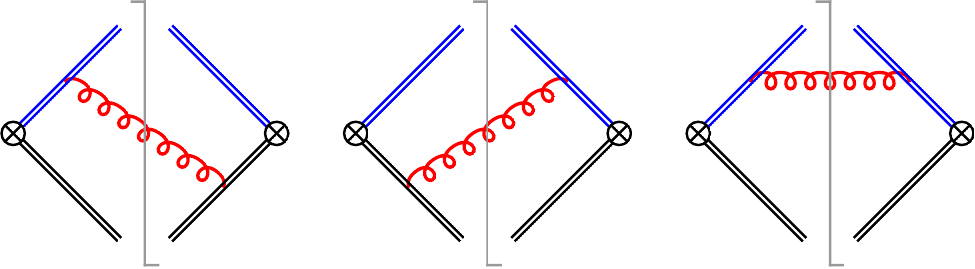}
    \color{blue}\put(10,74){$v_J^\mu$}
    \color{black}\put(8,8){$\bar n_J^\mu$}
    \end{overpic}
    \caption{One-loop Feynman diagrams of the collinear-soft function $S^c_{\mathcal{Q}}$. The blue and black lines indicate the Wilson lines along $v_J^\mu$ and $\bar n_J^\mu$ directions, respectively.}
    \label{fig:coftfunction}
\end{figure}

In this section, we calculate the one-loop perturbative expression for the collinear-soft function $S^c_{\mathcal{Q}}(\bm b,R,m_{\mathcal Q},\mu)$. The corresponding Feynman diagrams are shown in Fig.~\ref{fig:coftfunction}, where the blue and black lines represent Wilson lines along $v_J^\mu$ and $\bar n_J^\mu$ directions, respectively. Here, the massive quark velocity $v_J^\mu$ is defined by
\begin{align}
    v_J^\mu = \frac{\omega_J}{m_{\mathcal{Q}}} \frac{n_J^\mu}{2} + \frac{m_{\mathcal{Q}}}{\omega_J} \frac{\bar n_J^\mu}{2}\,,\quad\quad {\rm with}~v_J^2=1\,.
\end{align} 
Explicitly, the bare NLO collinear-soft function is given by 
\begin{align}
    S_{\mathcal{Q},{\rm NLO}}^{c}(\bm b,R,m_{\mathcal{Q}},\epsilon) = 2C_F \, w_{\bar n_J v_J}  - C_F \, w_{v_J v_J} \,,
\end{align}
where the collinear-soft integrals $w_{\alpha\beta}$ are defined in $b$-space as
\begin{align}
     w_{\alpha\beta} = \frac{\alpha_s \mu^{2\epsilon} \pi^\epsilon e^{\epsilon\gamma_E}}{2\pi^2} \int d^d k\, \delta^+(k^2) e^{- i \,\bar n_J \cdot k \,n_J \cdot b /2}
     \frac{\alpha \cdot \beta}{\left(\alpha\cdot k\right) \left(\beta\cdot k\right)} \, \theta\left[ \frac{n_J \cdot k  }{\bar n_J \cdot k } - \left( \frac{R}{2\cosh y_J} \right)^2\right]\,.
\end{align}
Upon performing the $k$-integration, we obtain the following  expressions for $w_{\alpha\beta}$

\begin{align}
    \label{eq:softints}
    w_{\bar n_J v_J} = & \frac{\alpha_s}{4\pi} \left[ -\frac{1}{\epsilon^2} - \frac{1}{\epsilon} \left(\ln \frac{\mu^2}{\mu_b^2} + 2 \ln \frac{-2 i c_{bJ}}{R} - \ln \frac{m_{\mathcal{Q}}^2 + p_T^2R^2}{p_T^2R^2}\right)\right]  +w_{\bar n_J v_J}^{\mathrm{fin}}  \,,\\
    w_{v_J v_J} =& \frac{\alpha_s}{4\pi}  \left[-\frac{1}{\epsilon}\left(\frac{2m_{\mathcal{Q}}^2}{m_{\mathcal{Q}}^2 + p_T^2 R^2}\right) \right] + w_{v_J v_J}^{\mathrm{fin}}\,.
\end{align}
We see that the finite quark mass corrections only enter into the single pole structure of the collinear-soft function. This is analogous to the observation made in Sec.~\ref{sec:jetfunction} in analyzing the massive quark jet function, and can be understood through the same physical reasoning. The finite terms are given by
\begin{align}
     w_{\bar n_J v_J}^{\mathrm{fin}}& = \frac{\alpha_s}{4\pi}\Biggl[2\left(\ln\frac{m_{\mathcal{Q}}^2+p_T^2R^2}{p_T^2 R^2}-\ln\frac{-2i\,c_{bJ}}{R}\right)\ln\frac{-2i\,c_{bJ}}{R} \\
    &\hspace{-0.3cm}+\left(\ln\frac{m_{\mathcal{Q}}^2+p_T^2R^2}{p_T^2R^2}-2\ln\frac{-2i\,c_{bJ}}{R}-\frac{1}{2}\ln\frac{\mu^2}{\mu_b^2}\right)\ln\frac{\mu^2}{\mu_b^2} +\mathrm{Li}_2\left(-\frac{m_{\mathcal{Q}}^2}{p_T^2R^2}\right)-\frac{\pi^2}{4}\Biggr]\,, \notag \\
    w_{v_J v_J}^{\mathrm{fin}}& = \frac{\alpha_s}{4\pi}\left[2\ln\frac{m_{\mathcal{Q}}^2+p_T^2R^2}{p_T^2R^2}-\frac{2m_{\mathcal{Q}}^2}{m_{\mathcal{Q}}^2+p_T^2R^2}\left(2\ln\frac{-2i\,c_{bJ}}{R}+\ln\frac{\mu^2}{\mu_b^2}\right)\right]\,.
\end{align}
Here, we note that $w_{\bar n_J v_J}$ reduces to the massless $w_{\bar n_J n_J}$ function \cite{Buffing:2018ggv,Chien:2019gyf} as $m_{\mathcal{Q}}\to 0$, while $w_{v_J v_J}$ vanishes. Given the expression for $S_{\mathcal{Q}}^c$, we can calculate the renormalization constant $Z^{S_{\mathcal{Q}}^{c}}$, which is given by
\begin{align}
    Z^{S_{\mathcal{Q}}^{c}} & = 1 \\
    &\hspace{-0.2cm}+ \frac{\alpha_s}{4\pi} C_F \left[ -\frac{2}{\epsilon^2} - \frac{2}{\epsilon} \left(\ln \frac{\mu^2}{\mu_b^2} + 2 \ln \frac{-2 i c_{bJ}}{R} - \frac{m_{\mathcal{Q}}^2}{m_{\mathcal{Q}}^2 + p_T^2 R^2} - \ln \frac{m_{\mathcal{Q}}^2 + p_T^2R^2}{p_T^2R^2} \right) \right]\,. \notag
\end{align}
This renormalization constant leads to the following formula for global collinear-soft anomalous dimension
\begin{align}\label{eq:cs-ad}
    \Gamma^{cs_{\mathcal{Q}}}(\alpha_s) = C_F \gamma^{\rm cusp}(\alpha_s) \ln \frac{R^2 \mu_b^2}{\mu^2} + \gamma^{cs_{\mathcal{Q}}}(\alpha_s)\,,
\end{align}
where the one-loop single logarithmic anomalous dimension is
\begin{align}
    \gamma_0^{cs_{\mathcal{Q}}}= -4 C_F \left[2 \ln \left(-2i c_{bJ}\right) - \frac{m_{\mathcal{Q}}^2}{m_{\mathcal{Q}}^2+p_T^2R^2} - \ln\frac{m_{\mathcal{Q}}^2 + p_T^2R^2}{p_T^2R^2}\right]\,.
\end{align}
The anomalous dimension for the collinear-soft function associated with the anti-quark is given by
\begin{align}
    \Gamma^{cs_{\bar{\mathcal{Q}}}}(\alpha_s) = \Gamma^{cs_{\mathcal{Q}}}(\alpha_s)|_{\phi_J \rightarrow \phi_J+\pi}\,.
\end{align}

For phenomenological purposes, we utilize the $\phi_b$-averaged collinear-soft function, which can be obtained through Eq.~\eqref{eq:softints} by making use of the following integrals
\begin{align}
    \int_0^{2\pi}\frac{d\phi_b}{2\pi} \ln\left(-2i c_{bJ}\right) = 0\,,
    \qquad
    \int_0^{2\pi}\frac{d\phi_b}{2\pi} \ln^2\left(-2i c_{bJ}\right) = -\frac{\pi^2}{6}\,.
\end{align}
The resulting anomalous dimension for the $\phi_b$-averaged collinear-soft function is denoted by $\bar{\Gamma}^{cs_{\mathcal{Q}}}$ with
\begin{align}
    \bar{\gamma}_0^{cs_{\mathcal{Q}}}= 4 C_F \left(  \frac{m_{\mathcal{Q}}^2}{m_{\mathcal{Q}}^2+p_T^2R^2} + \ln\frac{m_{\mathcal{Q}}^2 + p_T^2R^2}{p_T^2R^2}\right)\,.
\end{align}
Upon integrating over $\phi_b$, we find that $\bar{\Gamma}^{cs_{\bar{\mathcal{Q}}}}(\alpha_s) = \bar{\Gamma}^{cs_{\mathcal{Q}}}(\alpha_s)$ and, therefore, the two averaged collinear-soft functions $\bar S_{\mathcal Q}^c$ and $\bar S_{\bar{\mathcal Q}}^c$ behave identically under QCD evolution.

\subsection{Renormalization group consistency}\label{sec:rge}
Armed with the anomalous dimensions of each component, we are now positioned to demonstrate the RG consistency of our factorization framework.

Inspection of Eqs. \eqref{eq:jf-ad} and \eqref{eq:cs-ad} reveals that all mass corrections cancel exactly in the sum $\Gamma^{j_{\mathcal{Q}}}+\Gamma^{cs_{\mathcal{Q}}}$, making the RG consistency of our formalism identical to the massless case.
A similar observation is made in \cite{Kim:2020dgu}. This general physical behavior has also been observed in the context of inclusive HF jet production \cite{Dai:2018ywt}, where the authors offer the intuitive argument that as the heavy quark mass $m_{\mathcal{Q}}$ constitutes IR information, it thus does not affect the UV behavior of the semi-inclusive jet function. In the present context, we see that the UV evolution behavior of the product of the jet and collinear-soft functions is insensitive to the IR scale introduced by the heavy quark mass. However, in Sec. \ref{sec:xsec}, we will see how the heavy quark mass enters non-trivially and crucially into the evaluation of the differential cross section.

Therefore, upon combining Eqs. \eqref{eq:hard-ad}, \eqref{eq:beam-ad}, \eqref{eq:soft-ad},  \eqref{eq:jet-ad} and \eqref{eq:cs-ad}, the RG consistency of our formalism is established:
\begin{align}\label{eq:rge1}
    \Gamma^h + \Gamma^s +\Gamma^{f_g} + 2\Gamma^{j_{\mathcal{Q}}} + \Gamma^{cs_{\mathcal{Q}}}+ \Gamma^{cs_{\bar{\mathcal{Q}}}} =0\,.
\end{align}
Furthermore, we note that this consistency is preserved under the operation of $\phi_b$-averaging
\begin{align}\label{eq:rge2}
    \Gamma^h + \bar{\Gamma}^s +\Gamma^{f_g} + 2\Gamma^{j_{\mathcal{Q}}} + \bar{\Gamma}^{cs_{\mathcal{Q}}}+ \bar{\Gamma}^{cs_{\bar{\mathcal{Q}}}} =0\,.
\end{align}

\subsection{Resummation formula}\label{sec:xsec}
Utilizing our EFT framework, all-order resummation is achieved through RG evolution. The resulting all-order expression for the HF dijet production cross section is given at NLL\footnote{In our framework, we ignore contributions from NGL resummation. Such resummation could be included multiplicatively by using the parton shower algorithm developed recently for massive particles \cite{Balsiger:2020ogy}. Note that the fitting function used in \cite{Chien:2019gyf,Kang:2020xez} to capture the effects of NGLs is only an approximation for HF jet production, as finite heavy quark mass corrections are not included. } by 
\begin{align}\label{eq:res-unp}
    & \frac{d\sigma^{UU}}{dQ^2dyd^2\bm q_T dy_J d^2\bm p_T} = H(Q,y,p_T,y_J,\mu_h) \int_0^\infty \frac{b db}{2\pi}J_0(b\,q_T) f_{g/N}(x,\mu_{b*}) \notag \\
    &\hspace{0.25in} \times \exp \left[-\int_{\mu_{b *}}^{\mu_{h}} \frac{d \mu}{\mu} \Gamma^h\left(\alpha_{s}\right)-2\int_{\mu_{b *}}^{\mu_{j}} \frac{d \mu}{\mu} \Gamma^{j_{\mathcal{Q}}}\left(\alpha_{s}\right)-\int_{\mu_{b_{*}}}^{\mu_{c s}} \frac{d \mu}{\mu} \left(\bar\Gamma^{c s_{\mathcal{Q}}}\left(\alpha_{s}\right)+\bar \Gamma^{c s_{\bar{\mathcal{Q}}}}\left(\alpha_{s}\right)\right)\right]\notag \\
    &\hspace{0.25in} \times \exp\left[-S_{\rm NP}(b,Q_0,n\cdot p_g)\right]\,,
\end{align}
where $J_0$ is the zeroth order Bessel function of the first kind. In this expression, $\mu_h$, $\mu_j$, and $\mu_{cs}$ are the hard, jet, and collinear-soft scales, respectively. We have also performed the usual operator product expansion (OPE) of the unpolarized gluon TMD PDF $f_{g/N}(x,b,\mu, \zeta)$ in terms of the collinear gluon PDF $f_{g/N}(x,\mu)$ at the initial scales $\zeta_i = \mu_i^2 = \mu_{b*}^2$, and have kept the coefficient at LO to be consistent with NLL accuracy. The matching coefficient at higher-orders can be found in e.g.~\cite{Echevarria:2015uaa,Catani:2011kr,Gehrmann:2014yya,Luebbert:2016itl,Echevarria:2016scs,Luo:2019bmw,Ebert:2020yqt}. The function $S_{\rm NP}$ parameterizes the contribution from non-perturbative power corrections which are enhanced for $q_T\sim \Lambda_{\rm QCD}$. Explicitly, we apply the formula given in \cite{Su:2014wpa}, which reads
\begin{align}\label{eq:S_NP}
    S_{\mathrm{NP}}\left(b, Q_{0}, n\cdot p_g\right)=g_{1} b^{2}+\frac{g_{2}}{2} \frac{C_A}{C_{F}} \ln \frac{n\cdot p_g}{Q_{0}} \ln \frac{b}{b_{*}}\,.
\end{align}
We assign the following values to the parameters: $g_{1}=0.106\,\mathrm{GeV}^2$, $g_{2}=0.84$ and $ Q_{0}^{2}=2.4 \, \mathrm{GeV}^{2}$, and use the standard $b_*$-prescription \cite{Collins:1984kg}
\begin{align}
b_* = \frac{b}{\sqrt{1+b^2/b_{\rm max}^2}},\quad\qquad {\rm with}~~ b_{\rm max} = 1.5 {\rm ~GeV}^{-1}
\end{align}
in order to regularize the Landau singularity as $b\to \infty$. 

Moreover, the spin-dependent cross section is expressed as
\begin{align}\label{eq:res-siv}
    & \frac{d\sigma^{UT}(\bm S_T)}{dQ^2dyd^2\bm q_T dy_J d^2\bm p_T} = \sin(\phi_q-\phi_s )\, H(Q,y,p_T,y_J,\mu_h)\, \int_0^\infty \frac{b^2 db}{4\pi}J_1(b\,q_T) f_{1T, g/N}^{\perp, f}(x,\mu_{b*}) \notag \\
    &\hspace{0.25in}\times \exp \left[-\int_{\mu_{b *}}^{\mu_{h}} \frac{d \mu}{\mu} \Gamma^h\left(\alpha_{s}\right)-2\int_{\mu_{b *}}^{\mu_{j}} \frac{d \mu}{\mu} \Gamma^{j_{\mathcal{Q}}}\left(\alpha_{s}\right)-\int_{\mu_{b_{*}}}^{\mu_{c s}} \frac{d \mu}{\mu} \left(\bar\Gamma^{c s_{\mathcal{Q}}}\left(\alpha_{s}\right)+\bar\Gamma^{c s_{\bar{\mathcal{Q}}}}\left(\alpha_{s}\right)\right)\right]\notag \\
    &\hspace{0.25in}\times \exp\left[-S^\perp_{\rm NP}(b,Q_0,n\cdot p_g)\right]\,.
\end{align}
Here, we have expected a similar OPE for the gluon Sivers function $f_{1T, g/N}^{\perp}(x, b, \mu, \zeta)$ at the initial scales $\zeta_i = \mu_i^2 = \mu_{b*}^2$ and simply expressed the corresponding collinear function at LO as $f_{1T, g/N}^{\perp, f}(x,\mu_{b*})$ for simplicity. In principle, the corresponding collinear functions in the OPE expansion would be the twist-3 three-gluon correlation functions defined in~\cite{Ji:1992eu,Beppu:2010qn}. To the best of our knowledge, detailed OPE calculations for the corresponding coefficient functions are not available in the literature. An expansion of the gluon Sivers function in terms of the collinear twist-3 quark-gluon-quark correlator, or the so-called Qiu-Sterman function~\cite{Qiu:1991pp,Qiu:1991wg}, in transverse momentum space is performed in~\cite{Yuan:2008vn}. On the other hand, the coefficient functions for the expansion of the quark Sivers function in terms of the three-gluon correlation functions are provided in~\cite{Dai:2014ala,Scimemi:2019gge}. The computation of the coefficient functions for expanding the gluon Sivers function in terms of the three-gluon correlation functions is essential for a full understanding of the QCD evolution of the gluon Sivers function. We leave this to future work. 

Our knowledge about gluon Sivers functions, especially in the proper TMD factorization formalism, is rather limited. At the present moment, the only experimental constraint on the gluon Sivers function, in the TMD framework, comes from the SIDIS measurement of back-to-back hadron pairs
off transversely-polarized deuterons and protons at COMPASS~\cite{Adolph:2017pgv}. However, as of yet, there has been no theoretical extraction of the gluon Sivers function from such data. On the other hand, an important theoretical constraint on the gluon Sivers function comes from the Burkardt sum rule~\cite{Burkardt:2004ur}. For the phenomenological purposes of the next section, we adopt the non-perturbative parameterization utilized by \cite{DAlesio:2015fwo,Aschenauer:2015ndk}~\footnote{Note that the gluon Sivers function in~\cite{DAlesio:2015fwo}, and its updated version~\cite{DAlesio:2018rnv}, is constrained to their study of the $p^\uparrow p \to \pi X$ process. Technically, this is not subject to a TMD factorization framework, but it serves as a starting point for our numerical study, following \cite{Zheng:2018ssm}.}. Specifically, for the non-perturbative Sudakov, we take 
\begin{align}\label{eq:NPSudakov}
    S_{\mathrm{NP}}^{\perp}\left(b, Q_{0}, n\cdot p_g\right)=g_{1}\rho\, b^{2}+\frac{g_{2}}{2} \frac{C_A}{C_F}\ln \frac{n\cdot p_g}{Q_{0}} \ln \frac{b}{b_{*}}\,, 
\end{align}
where the $g_2$-dependent term is spin-independent and is, therefore, the same term occurring in Eq.~\eqref{eq:S_NP}, while the term $\propto g_1\rho$ can be connected to the Gaussian width in transverse momentum space~\cite{Echevarria:2014xaa} for the gluon Sivers function. For the collinear part of the gluon Sivers function, $f_{1T, g/p}^{\perp, f}(x,\mu)$ in Eq.~\eqref{eq:res-siv}, we take
\begin{align}
     f_{1T,g/N}^{\perp f}(x,\mu) = N_g \frac{4\rho\sqrt{2e\rho(1-\rho)g_1}}{M_{\rm proton}}x^{\alpha_g}(1-x)^{\beta_g}\frac{(\alpha_g+\beta_g)^{\alpha_g+\beta_g}}{\alpha_g^{\alpha_g}\beta_g^{\beta_g}}f_{g/N}(x,\mu)\,,
\end{align}
with the parameters given by
\begin{align}
    N_g = 0.65,\quad \alpha_g=2.8,\quad \beta_g=2.8,\quad \rho=0.5,\quad M_{\rm proton}=1\,{\rm GeV}\,,
\end{align}
and $f_{g/N}(x,\mu)$ denoting the unpolarized collinear gluon PDF. For $f_{g/N}(x,\mu)$, we use CT14nlo \cite{Dulat:2015mca}--specifically, CT14nlo\_NF3 (CT14nlo\_NF4) for charm (bottom) jet-pair production with 3 (4) active parton flavors. 

At this point, it is important to note that while the mass corrections in sum of the anomalous dimensions for the collinear-soft and massive jet functions cancel, the mass-dependence of $\Gamma^{j_{\mathcal{Q}}}$ contributes to the differential cross section. By examining Eqs.~\eqref{eq:res-unp} and \eqref{eq:res-siv}, we see that the mass corrections enter into the evolution between the scales $\mu_j$ and $\mu_{cs}$. We will see in the following section that this can significantly affect both the $q_T$-distributions and spin asymmetries for HF dijet production at the EIC.

\section{Numerical results}\label{sec:num}
In this section, we present numerical results for HF dijet production in unpolarized and transversely-polarized-proton-electron collisions at the future EIC. We set the energies of the electron and proton beam to be 20 GeV and 250 GeV, respectively. These beam-energy values yield a electron-proton center-of-mass energy of $\sqrt{S_{\ell P}} = 141$~GeV. For the all-order resummation formulae in Eqs.~\eqref{eq:res-unp} and \eqref{eq:res-siv}, the renormalization scales for each function are chosen to be
\begin{align}\label{eq:scale}
    \mu_h = \sqrt{Q^2+p_T^2},\quad \mu_j=p_T R,\quad \mu_{cs}=\mu_{b_*}R\,.
\end{align}
Here, note that the Landau singularity associated with the collinear-soft scale is also regularized by the $b_*$-prescription. 
\begin{figure}[t]
\begin{center}
\includegraphics[width=0.44\textwidth]{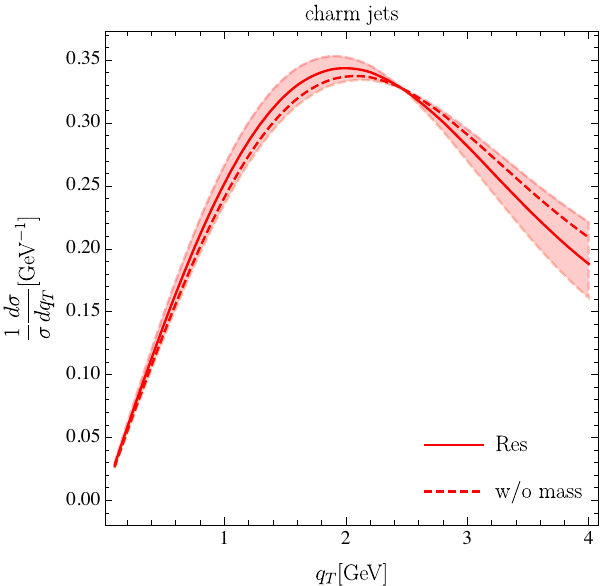}\hspace{0.5cm}
  \includegraphics[width=0.44\textwidth]{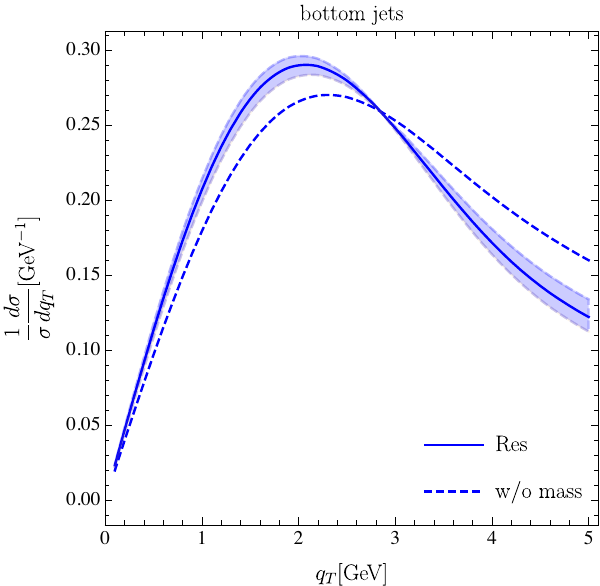}
\end{center}
  \caption{The normalized $q_T$-distribution for the unpolarized cross section of charm (left plot) and bottom (right plot) dijet production at the EIC. The solid curves are the results from using the resummation formula Eq.~\eqref{eq:res-unp}, while the dashed curves represent the resummation prediction using the evolution kernel without finite quark mass corrections. The red and blue bands indicate theoretical uncertainties from the variation of hard and jet scales as discussed in the text.} 
\label{fig:pheno}
 \end{figure} 

As given in the calculation of the jet function, we consider HF jets constructed using the anti-$k_T$ algorithm with radius $R=0.6$. The corresponding kinematic cuts for charm and bottom jets in the Breit frame are 
\begin{align}\label{eq:cb-ranges}
    {\rm charm~jets:}&~~5\,{\rm GeV}<p_T<10\,{\rm GeV},\quad |y_J|<4.5\,,\notag \\
    {\rm bottom~jets:}&~~10\,{\rm GeV}<p_T<15\,{\rm GeV},\quad |y_J|<4.5\,,
\end{align}
respectively. The charm and bottom quark masses are chosen as $m_c=1.5\,{\rm GeV}$ and $m_b=5\,{\rm GeV}$. The spin asymmetry from the gluon Sivers function is defined as
\begin{align}
    A_{UT}^{\sin(\phi_q-\phi_s)} = 2 \frac{\int d\phi_s d\phi_q \sin(\phi_q-\phi_s) \, d\sigma^{UT}(\bm S_T)}{\int d\phi_s d\phi_q \, d\sigma^{UU} }\,.
\end{align}

\begin{figure}[t]
\begin{center}
\includegraphics[width=0.44\textwidth]{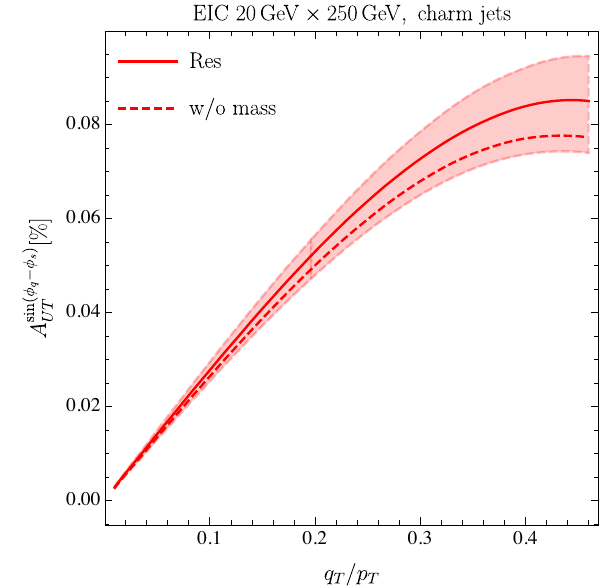}\hspace{0.5cm}
  \includegraphics[width=0.44\textwidth]{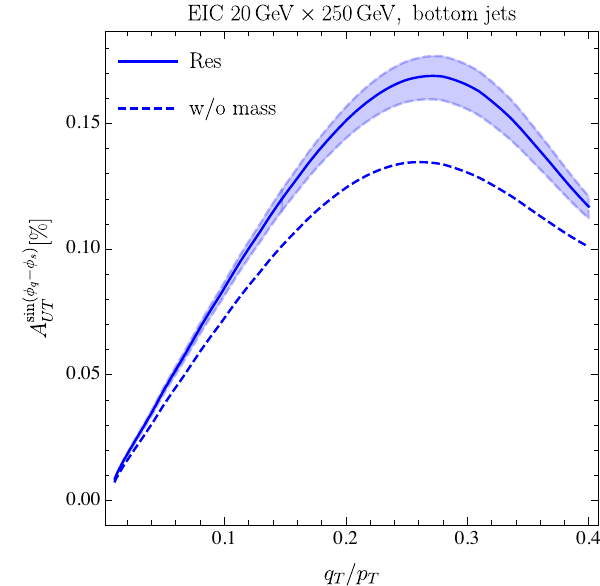}
\end{center}
  \caption{The Sivers spin asymmetry for charm (left plot) and bottom (right plot) dijet production at the EIC is plotted as a function of $q_T/p_T$. The solid curves are the results from using the resummation formula, while the dashed curves represent the resummation prediction using the evolution kernel without finite quark mass corrections. The red and blue bands indicate theoretical uncertainties from the variation of hard and jet scales. }
\label{fig:siv}
 \end{figure}

In Fig.~\ref{fig:pheno}, we display the normalized unpolarized cross section, $1/\sigma\, d\sigma/dq_T$, as a function of the imbalance $q_T$. In Fig.~\ref{fig:siv}, the Sivers spin asymmetry $A_{UT}^{\sin(\phi_q-\phi_s)}$ is presented as a function of $q_T/p_T$ following~\cite{Arratia:2020nxw}, for both charm (left panel) and bottom (right panel) jets, respectively. For both plots, the solid curves are the results obtained using the resummation formula, while the dashed curves represent the resummation prediction using the evolution kernel without finite quark mass corrections. For both the unpolarized $q_T$ and $A_{UT}^{\sin(\phi_q-\phi_s)}$ distributions, we find that the effects of the finite quark masses are modest for charm jets and quite sizable for bottom jets. This can be attributed to the sizes of the charm and bottom masses relative to their associated jet scales $p_TR$. As discussed in Secs.~\ref{sec:jetfunction} and \ref{sec:coftfunction}, we have that $J_{\mathcal{Q}}\rightarrow J_q$ and $S^c_{\mathcal{Q}}\rightarrow S^c_q$ as $m_{\mathcal{Q}}\rightarrow 0$, making them analytic functions of $m_{\mathcal{Q}}$ in the neighborhood of zero mass. Since Eqs.~\eqref{eq:res-unp} and~\eqref{eq:res-siv} carry their mass-dependence through the anomalous dimensions for the jet and collinear-soft functions, Eqs.~\eqref{eq:jf-ad} and~\eqref{eq:cs-ad}, one sees that the massive versions of these functions are connected to the massless versions by the ratio $m_{\mathcal{Q}}/\left(p_TR\right)$--it is in fact this dimensionless parameter that controls the physical size of the mass corrections. With this in mind, one sees that Eq.~\eqref{eq:cb-ranges} naturally positions bottom dijets further (in terms of the parameter $m_{\mathcal{Q}}/\left(p_TR\right)$) from light flavor jet-pairs than it does charm dijets. This relative positioning is then clearly displayed in Figs.~\ref{fig:pheno} and~\ref{fig:siv}. 

In order to estimate the theoretical uncertainties, in both Figs.~\ref{fig:pheno} and~\ref{fig:siv} we also show the uncertainties from scale variations, which are given by the red and blue bands. Here we vary the hard and jet scales by a factor of two around their default values as defined in \eqref{eq:scale}, and the total uncertainty bands are obtained by the envelope of all the variations.  Since the non-perturbative Sudakov factor in Eq. \eqref{eq:NPSudakov} is fitted  at the canonical scale $\mu_{b_*}$, we do not include theory uncertainties from $\mu_{b_*}$ and $\mu_{cs}$ variations. We find that the scale uncertainty is compatible with the finite quark mass corrections in charm dijet process, while its impact on the bottom dijet process is smaller than the mass correction.  Therefore in order to identify the finite quark mass effects in the charm dijet process it is essential to reduce the scale uncertainties. Our factorization and resummation formula provides a clear structure to improve the perturbative accuracy, which makes scale uncertainty further reduction possible. We leave the higher-order perturbative calculations in future work.

\section{Conclusion}\label{sec:concl}
A major priority of the future EIC is to explore the gluon TMD PDFs. In this paper, we have investigated the use of back-to-back HF dijet production in transversely-polarized target DIS as a means of  probing spin-dependent gluon TMD PDFs. We have calculated the expressions for the mass-dependent jet and collinear-soft functions at next-to-leading order. Using these expressions, as well as Soft-Collinear Effective Theory, we resum the large logarithms associated with these expressions at next-to-leading logarithmic accuracy. We then provide a factorization theorem for this process with QCD evolution in the kinematic region where heavy quark mass $m_{\mathcal{Q}} \lesssim p_T R \ll p_T$, with $p_T$ and $R$ being the transverse momentum and the radius of the jet, respectively. Furthermore, we generate a prediction for the Sivers asymmetry for charm and bottom dijets at the EIC, which can be used to probe the gluon Sivers function. We carefully study the effects of the HF masses by comparing our mass-dependent predicted asymmetry against the asymmetry in the massless limit. We find that, in the kinematic region we consider, the HF masses generate modest corrections to the predicted asymmetry for charm dijet production but sizable corrections for the bottom dijet process. Furthermore, we also consider the theoretical uncertainties from the scale variation. We find that the scale uncertainty can be compatible with the corrections from finite quark mass effects, especially for charm dijets production. In order to identify the mass effects and reduce the scale uncertainties one has to include higher-order corrections in the matching coefficients and the corresponding anomalous dimensions in Eqs. \eqref{eq:res-unp} and \eqref{eq:res-siv},  and we leave the detailed perturbative calculations in future work. 

\section*{Acknowledgements}
We thank Miguel Arratia and Kyle Lee for helpful discussions. Z.K. and D.Y.S. are supported by the National Science Foundation under CAREER award PHY-1945471. J.R. is supported by the UC Office of the President through the UC Laboratory Fees Research Program under Grant No.~LGF-19-601097. J.T. is supported by NSF Graduate  Research Fellowship Program under Grant No. DGE-1650604. D.Y.S. is also supported by Center for Frontiers in Nuclear Science of Stony Brook University and Brookhaven National Laboratory. This work is supported within the framework of the TMD Topical Collaboration.

\bibliographystyle{JHEP}
\bibliography{jet.bib}

\end{document}